\documentclass{revtex4}
\usepackage{graphicx}
\usepackage{fancyhdr}
\usepackage{amsmath}
\usepackage{color}
\usepackage{url}
\usepackage{amssymb}

\definecolor{orange}{rgb}{1, 0.4, 0} 
\definecolor{vertfonce}{rgb}{0, 0.4, 0} 
\definecolor{marron}{rgb}{0.36,0.13,0.00} 
\definecolor{purple}{rgb}{0.4,0.0,0.4} 
\definecolor{pink}{rgb}{0.8,0.3,0.6} 
\definecolor{gray}{rgb}{0.3,0.3,0.3} 

\def\csq{\chi^2}

\def\kg{\kappa_g}
\def\kgam{\kappa_\gamma}

\def\csqk{\chi^2(\kappa)}
\def\hmu{\hat\mu}
\def\muik{\mu_i(\kappa)}
\def\muk{\mu(\kappa)}
\def\zetaip{\zeta_i^p}
\def\csqkp{\chi^{2\prime}(\kappa)}
\def\mug{\mu_{\text{ggH/ttH}}}
\def\muv{\mu_{VBF/VH}}
\def\hmug{\hat{\mu}_{\text{ggH/ttH}}}
\def\hmuv{\hat{\mu}_{VBF/VH}}

\begin{document}

\title{Higgs couplings beyond the Standard Model}

\author{Guillaume Drieu La Rochelle}
\affiliation{CNRS/IN2P3, UMR5822, Institut de Physique Nucl\'eaire de Lyon, F-69622 Villeurbanne 
Cedex, France}

\begin{abstract}
We consider the Higgs boson decay processes and its production and provide a parameterisation 
tailored for testing models of new physics. The choice of a particular 
parameterisation depends on a non-obvious balance of quantity and quality of the available 
experimental data, envisaged purpose for the parameterisation and degree of model independence. At 
present only simple parameterisations with a limited number of fit parameters can be performed, but 
this situation will improve with the forthcoming experimental LHC data. It is therefore important 
that different approaches are considered and that the most detailed information is made available to 
allow testing the different aspects of the Higgs boson physics and the possible hints beyond the 
Standard Model.
\end{abstract}

\maketitle

\thispagestyle{fancy}

\section{Higgs coupling parameterisations}

Since the discovery of the Higgs boson last year by both ATLAS and CMS collaboration, it is clear 
that the study of its couplings is likely to give valuable information on the nature 
of the electroweak symmetry breaking and eventually to probe New Physics scenarios. Indeed, many of 
the currently studied BSM (Beyond the Standard Model) models can generically lead to deviations in 
the 
Higgs couplings, allowing for a constraint complementary to the direct searches for extra 
particles. However, to achieve such a goal, one must first make contact between the theory side 
(that is, the large amount of BSM models) and the experimental one (which are the results presented 
by the collaborations), and this implies the choice of a parameterisation.

From the experimental point of view it makes sense to just parameterise Higgs physics in terms of 
observed quantities such as 
branching ratios and cross-sections. This is for example the case of the parameterisation proposed 
in 
Ref.~\cite{LHCHWG_reco_2012}, where the relevant cross-sections and partial decay widths are 
multiplied by a suitable factor. The advantage of such an approach is its simple link to the 
experimentally 
measured quantities. On the other hand, with such a choice, correlations among the different 
parameters are not explicit, in particular between tree level and loop induced observables. For 
example, a modification of the couplings to $W$ bosons and top will also affect the loop-level 
couplings for the Higgs production via the 
gluon channel or the Higgs decay into two photons. Instead, we propose thus in 
\cite{gdlr_higgs_2012} a parameterisation where the contribution of loops of New Physics to the $H 
\to g g$
and $H \to \gamma \gamma$ modes is explicitly disentangled from the modification of tree level
couplings.

We must point that out that many studies have also been carried out relying on an effective field 
theory approach 
(\cite{bonnet_higgs_2011,bonnet_higgs_2011,corbett_higgs_2012,grojean_higgs_1202,grojean_higgs_1207}
). While this approach has the important 
feature of allowing a full treatment of the radiative corrections (see 
\cite{passarino_higgs_2012}), 
it often relies on a large set of parameters, for which the current experimental accuracy is still 
lacking a bit behind. A similar study has also been recently carried 
out in \cite{belanger_higgs_2012}. 

\section{Using experimental Higgs results}
In order to use the latest data from ATLAS and CMS collaboration 
(\cite{atlas_comb_hcp_2012,cms_comb_hcp_2012}), one has to define the compatibility of a given 
model 
with data. The first 
approach is to use a $\csq$ test based on the signal strength $\mu$ which is 
simply the 
cross-section for $pp\to H\to XX$ normalised to the Standard Model expectation. Thus the $\csq$ 
reads
\begin{equation}
 \csqk=\sum_i\left(\frac{\hmu_i-\muik}{\sigma_i}\right)^2,
\label{eq:chi2_1}
\end{equation}
where $\hmu$ is the best fit reported by the experiment, $\muk$ the prediction of the model on the 
parameter point $\kappa$, $\sigma$ the uncertainty, and $i$ runs on all subchannels of each 
experiment. Then this $\csqk$ is compared to a standard $\csq$ distribution with $n$ degrees of 
freedom, where $n$ is the number of subchannels, in order to determine if the model is excluded or 
not. However this method suffers from a few shortcomings : first the $\muik$ does not correspond to 
the 
inclusive cross-section, but to the exclusive one. For instance the $H\to\gamma\gamma$ decay mode 
in 
ATLAS is divided into 11 subchannels, which amount to as many exclusive cross-sections. To compute 
exclusive cross-sections, one needs the experimental fractions per mode $\zetaip$, or equivalently 
the efficiencies :
\begin{equation}
 \mu_i^{XX}=\sum_p \zetaip \mu_p^{XX},
\end{equation}
where $\zetaip$ represent the SM fraction of the production mode $p$ among all production modes in 
the subchannel $i$, and $\mu_p^{XX}$ the signal strength of that specific production mode in the 
final 
state $XX$. Although the $\zetaip$ are computed by the collaborations themselves, there are not 
always publicly available. Another inconvenient raised by eq. \ref{eq:chi2_1} is that in adding all 
subchannels together we 
implicitly assume that they are not correlated. While this is the case of statistical uncertainty 
it 
is certainly not true for systematical uncertainty, let alone theory uncertainty.

\subsection{Improved $\csq$ method}
There exists nevertheless a way to improve the statistical test : indeed the collaborations have 
released 2D plots of the $\csq$ in each decay mode, showing the $\csq$ of this decay mode as a 
function of $\left(\mug,\muv\right)$. Here 
$\mug$ stands for a 
common signal strength for both $gg\to H$ and $pp\to \bar ttH$ and $\muv$ for both 
$VBF$ and $VH$. By approximating the $\csq$ to a gaussian, we can trade the 
simple $\csq$ test to a new one, defined as
\begin{equation}
 \csqkp=\sum_{XX}\left(\hmug^{XX}-\mug^{XX},\hmuv^{XX}-\muv^{XX}\right)\ V^{-1}\ 
\binom{\hmug^{XX}-\mug^{XX}}{\hmuv^{XX}-\muv^{XX}}
\end{equation}
where $V^{-1}$ is the inverse of the covariance matrix, deduced from the experimental plots. The 
advantages of the new method are straightforward : one does not need any knowledge of the fractions 
$\zetaip$ and moreover, most of the correlations between production modes are taken into account.\\

\subsection{Shortcomings of the improved method}
It is however clear that in order to use the improved $\csq$ one has to make some assumptions. 
Indeed 
we are collecting 4 productions modes ($gg\to H$, VBF, VH and $\bar ttH$) into 2 parameters 
($\mug$,$\muv$). However the requirement is not so stringent since any model abiding by custodial 
symmetry will feature an identical rescaling for both VBF and VH cross-sections. Concerning $gg\to 
H$ and $\bar ttH$, so far most channels are not sensitive to the latter because of its small 
cross-section, so assuming an identical rescaling does not affect much the prediction.

Another issue is that the $\csq$ reported by experiment is not a gaussian, and 
though this approximation may be accurate near the best fit, it goes worse as one moves away from 
it. However since there is no analytic form of the real $\csq$, there is no easy alternative to the 
gaussian approximation.

\subsection{Experimental input}
The experimental input consists in the best fits and covariance matrix for $\gamma\gamma$, $WW$ 
and $\bar\tau\tau$ decay modes for CMS and best fits and uncertainties for all subchannels in the 
remaining decay modes of CMS and all of ATLAS. The reason why we did not use improved $\csq$ for 
each decay mode is first that for decay mode $ZZ$ and $\bar bb$ there is no much gain since the 
former is an exclusive channel and the second relies entirely on associated vector boson 
production. Second, all data from ATLAS was not available at the time to carry out the 
full improved $\csq$ method. Those data were extracted 
\cite{cms_comb_hcp_2012,atlas_comb_hcp_2012}, 
and results for improved $\csq$ are shown in Table \ref{tab:exp}.

\begin{figure}[!h]
\begin{tabular}{|c|c|c|}
\hline
  Decay mode & $(\hmug,\hmuv)$ & $V$\\\hline
 $H\to\gamma\gamma$ & $(0.95.3.77)$ & $\begin{pmatrix}0.95&-1.35\\-1.35&6.87\end{pmatrix}$\\\hline
 $H\to WW$ & $(0.77.0.39)$ & $\begin{pmatrix}0.19&0.15\\0.15&1.79\end{pmatrix}$\\\hline
 $H\to\tau\tau$ & $(0.93.0.89)$ & $\begin{pmatrix}2.02&-0.92\\-0.92&2.14\end{pmatrix}$\\\hline
\end{tabular}
\caption{\label{tab:exp} {\em CMS results in $H\to\gamma\gamma$, $H\to WW$ and $H\to\tau\tau$ 
channels.}}
\end{figure}

\section{$\kg,\kgam$ parameterisation}
Our first parameterisation will be tailored to BSM models which alter mostly the Higgs physics 
via loop effects. This is a generic feature of models where there is no much mixing between 
the SM Higgs and any other scalars, but which feature extra particles light enough (light 
superpartners in Supersymmetry, light vector-like fermions in extra dimensional theories, and so 
on). In which case the effect of in each BSM set-up can be parameterised by the pair $(\kg,\kgam)$, 
that we define as the amplitude of new particles contributing to the partial 
widths $\Gamma_{gg}$ and $\Gamma_{\gamma\gamma}$ normalised by the SM top amplitude :
\begin{eqnarray} 
\Gamma_{\gamma\gamma}&=&\frac{G_F\alpha^2m_H^3}{128\sqrt{2}\pi^3}
\left|A_W(\tau_W)+C^\gamma_t3\left(\frac{2}{3}\right)^2A_t(\tau_t)(1+\kgam)+...\right|^2\\
\Gamma_{gg}&=&\frac{G_F\alpha_S^2m_H^3}{16\sqrt{2}\pi^3}
\left|C^g_t3\frac{1}{2}A_t(\tau_t)(1+\kg)+...\right|^2
\label{eq:Gamma}
\end{eqnarray}
where dots stands for the contribution of light quarks, $A_X(\tau_X)$ 
are usual SM amplitudes and $C_t^X$ contains the QCD NLO corrections (see \cite{gdlr_higgs_2012} 
for details).

\subsection{Results}
For reference, sample points for the following models are indicated:
\begin{itemize}
\item[-] [\textcolor{orange}{$\blacklozenge$}] fourth generation  where the result is independent 
on 
the masses and Yukawa couplings.
\item[-] [\textcolor{red}{$\ast$}] Littlest Higgs~\cite{LH}, where the result scales with the 
symmetry breaking scale $f$, set here to $f=500$ GeV for a model with $T$-parity.
\item[-] [\textcolor{vertfonce}{$\blacktriangle$}] Simplest Little Higgs~\cite{SLH}, where the 
result scales with the $W'$ mass, also set here to $m_{W'}=500$ 
GeV for a model with $T$-parity;
\item[-] [\textcolor{cyan}{$\blacksquare$}] colour octet model~\cite{octet}, where the result is 
inversely proportional to the mass $m_S = 750$ GeV in the example;
\item[-] [\textcolor{vertfonce}{$\otimes$}] 5D Universal Extra Dimension model~\cite{UED}, where 
only the top and $W$ resonances contribute and the result scales with the size of the extra 
dimension (here we set $m_{KK} = 500$ GeV);
\item[-] [\textcolor{blue}{$\bigstar$}] 6D UED model on the Real Projective Plane~\cite{RPP}, with 
$m_{KK}= 600$ GeV is set to the LHC bound~\cite{RPPb};
\item[-] [\textcolor{purple}{$\bullet$}] the Minimal Composite Higgs~\cite{GHUwarped} (Gauge Higgs 
unification in warped space) with the IR brane at $1/R' = 1$ TeV, where only $W$ and top towers 
contribute significantly and the point only depends on the overall scale of the KK masses;
\item[-] [\textcolor{gray}{$\blacktriangledown$}] a flat ($W'$ at 2 TeV) and 
[\textcolor{pink}{$\spadesuit$}] warped ($1/R'$ at 1 TeV) version of brane Higgs models. The 
result 
only depends on the overall scale of the KK masses.
\end{itemize}

\begin{figure}[tb]
\begin{center}
\includegraphics[width=0.45\textwidth]{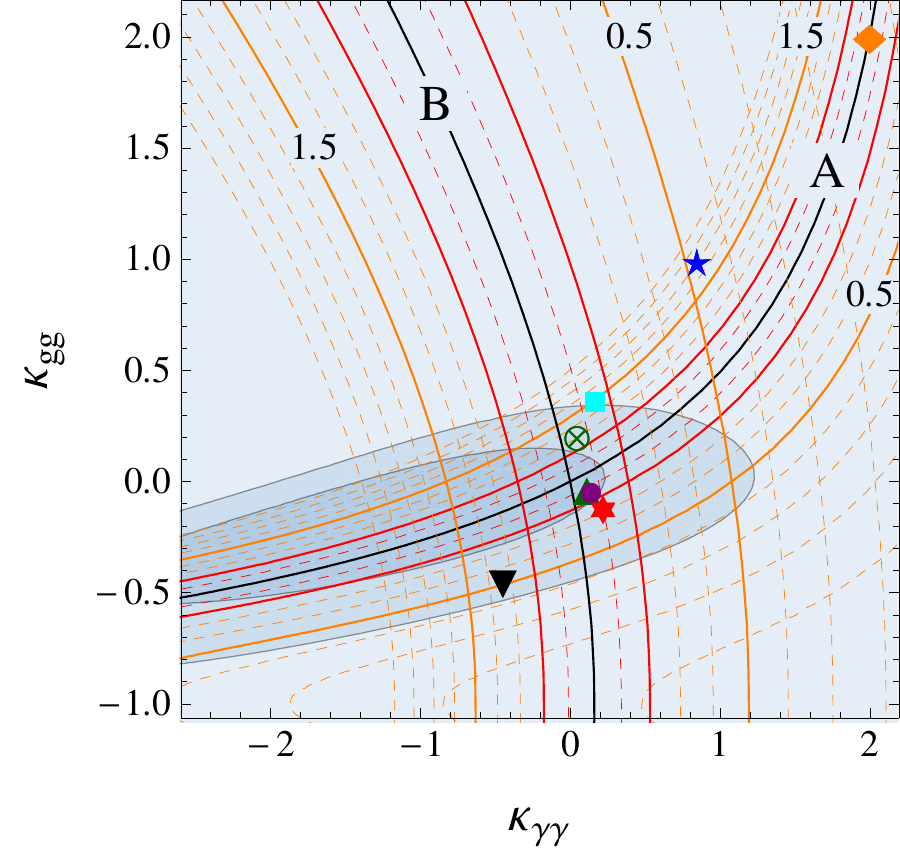}~~~~~
\includegraphics[width=0.45\textwidth]{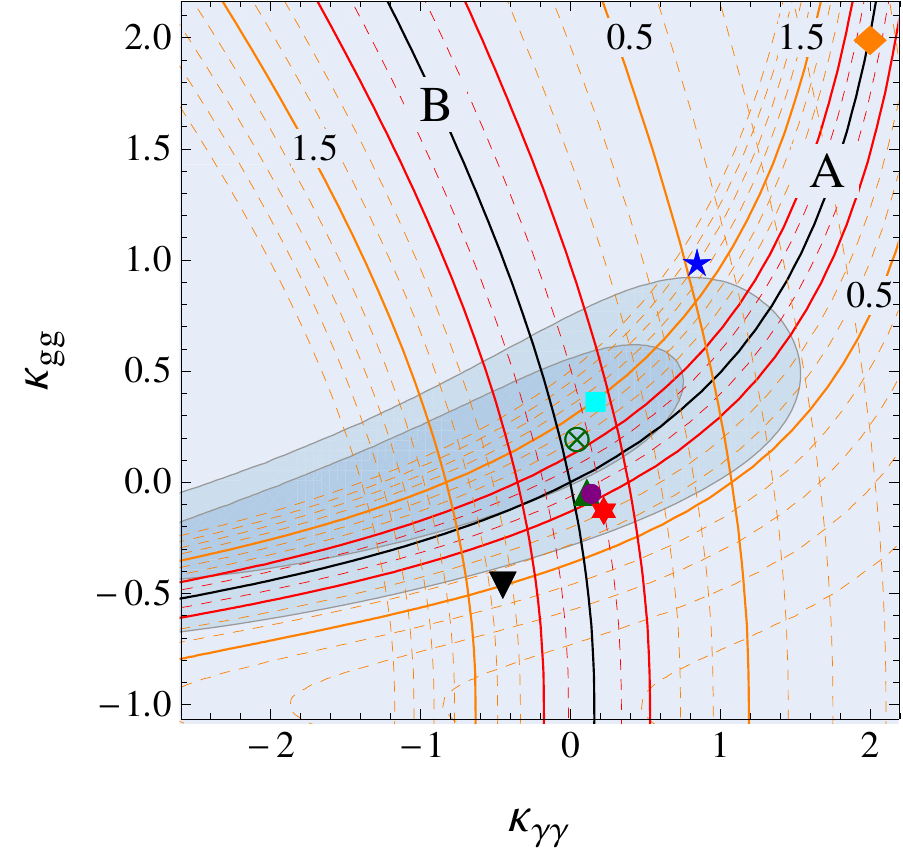}
\end{center}
\caption{\footnotesize  $\kappa'_{\gamma \gamma}$ and $\kappa'_{g g}$ at the LHC for a  Higgs boson 
with $m_H = 125$ GeV.
The two solid lines correspond to the SM values of the inclusive $\gamma \gamma$ channel ({\bf A}), 
and the vector boson fusion 
production channel ({\bf B}). On the left panel, the fit using ATLAS data. On the right, the fit
using CMS data. Both fits use $\gamma\gamma$, $ZZ$ and $\bar bb$ channels. Darker (lighter) blue
are 
the 1, 2 $\sigma$ limits.}
\label{fig:kgkgam2}
\end{figure}

One must note that each model will not be represented as a point in the $(\kg,\kgam)$ plane, but 
rather by a line starting at the SM point $(0,0)$, since they all have a decoupling limit, except 
for the $4^{th}$ generation. We show in figure \ref{fig:kgkgam2}, the one and two sigma excluded 
regions, and the position of 
the models with respect to those exclusions. As one can see, the $4^\text{th}$ generation 
lies well away from the compatible region, and so do some of the benchmark of the other models. In 
particular, the 6D UED benchmark that we used was chosen so that the heavy scale was at the limit 
of the direct searches for extra particles, and we see that in this case, the indirect bounds form 
Higgs physics does much better than the direct search.

\section{Conclusion}
We have shown how to go beyond the simplest methods when using experimental data to constrain 
Higgs couplings and account for part of the correlations between measurements. We have presented a 
parameterisation (\cite{llodra_higgs_2009,gdlr_higgs_2012}) and
we showed how it can be used for testing and putting exclusion limits on models of new physics 
beyond the Standard Model. In particular our parameterisation is tailored to investigate BSM 
models, keeping track of 
the specific correlations among the 
parameters. It also allows more easily to interpret mass limits and contributions to the loops 
giving the effective Higgs 
boson vertices. 

We also performed 2 parameter fits of the CMS and ATLAS results using all available channels, 
showing 
that they
already include all the necessary information and are therefore a good approximation at this stage. 
More precise measurements of extra channels will require the inclusion of more effective parameters.

\begin{acknowledgments}
I thank the organisers of the HPNP 2013 conference for this very pleasant conference. We 
thank L. Panizzi for discussions.
We also thank S. Shotkin-Gascon, N. Chanon and the CMS group in Lyon for useful discussions.

\end{acknowledgments}

\bigskip

\begin{thebibliography}{18}
\expandafter\ifx\csname natexlab\endcsname\relax\def\natexlab#1{#1}\fi
\expandafter\ifx\csname bibnamefont\endcsname\relax
  \def\bibnamefont#1{#1}\fi
\expandafter\ifx\csname bibfnamefont\endcsname\relax
  \def\bibfnamefont#1{#1}\fi
\expandafter\ifx\csname citenamefont\endcsname\relax
  \def\citenamefont#1{#1}\fi
\expandafter\ifx\csname url\endcsname\relax
  \def\url#1{\texttt{#1}}\fi
\expandafter\ifx\csname urlprefix\endcsname\relax\def\urlprefix{URL }\fi
\providecommand{\bibinfo}[2]{#2}
\providecommand{\eprint}[2][]{\url{#2}}

\bibitem[{\citenamefont{Group et~al.}(2012)\citenamefont{Group, A., A., M., M.
  et~al.}}]{LHCHWG_reco_2012}
\bibinfo{author}{\bibfnamefont{L.~H. C. S.~W.} \bibnamefont{Group}},
  \bibinfo{author}{\bibfnamefont{D.}~\bibnamefont{A.}},
  \bibinfo{author}{\bibfnamefont{D.}~\bibnamefont{A.}},
  \bibinfo{author}{\bibfnamefont{D.}~\bibnamefont{M.}},
  \bibinfo{author}{\bibfnamefont{G.}~\bibnamefont{M.}}, \bibnamefont{et~al.}
  (\bibinfo{year}{2012}), \eprint{1209.0040}.

\bibitem[{\citenamefont{Cacciapaglia et~al.}(2013)\citenamefont{Cacciapaglia,
  Deandrea, La~Rochelle, and Flament}}]{gdlr_higgs_2012}
\bibinfo{author}{\bibfnamefont{G.}~\bibnamefont{Cacciapaglia}},
  \bibinfo{author}{\bibfnamefont{A.}~\bibnamefont{Deandrea}},
  \bibinfo{author}{\bibfnamefont{G.~D.} \bibnamefont{La~Rochelle}},
  \bibnamefont{and} \bibinfo{author}{\bibfnamefont{J.-B.}
  \bibnamefont{Flament}}, \bibinfo{journal}{JHEP}
  \textbf{\bibinfo{volume}{1303}}, \bibinfo{pages}{029} (\bibinfo{year}{2013}),
  \eprint{1210.8120}.

\bibitem[{\citenamefont{Bonnet et~al.}(2012)\citenamefont{Bonnet, Gavela, Ota,
  and Winter}}]{bonnet_higgs_2011}
\bibinfo{author}{\bibfnamefont{F.}~\bibnamefont{Bonnet}},
  \bibinfo{author}{\bibfnamefont{M.}~\bibnamefont{Gavela}},
  \bibinfo{author}{\bibfnamefont{T.}~\bibnamefont{Ota}}, \bibnamefont{and}
  \bibinfo{author}{\bibfnamefont{W.}~\bibnamefont{Winter}},
  \bibinfo{journal}{Phys.Rev.} \textbf{\bibinfo{volume}{D85}},
  \bibinfo{pages}{035016} (\bibinfo{year}{2012}), \eprint{1105.5140}.

\bibitem[{\citenamefont{Corbett et~al.}(2012)\citenamefont{Corbett, Eboli,
  Gonzalez-Fraile, and Gonzalez-Garcia}}]{corbett_higgs_2012}
\bibinfo{author}{\bibfnamefont{T.}~\bibnamefont{Corbett}},
  \bibinfo{author}{\bibfnamefont{O.}~\bibnamefont{Eboli}},
  \bibinfo{author}{\bibfnamefont{J.}~\bibnamefont{Gonzalez-Fraile}},
  \bibnamefont{and}
  \bibinfo{author}{\bibfnamefont{M.}~\bibnamefont{Gonzalez-Garcia}},
  \bibinfo{journal}{Phys.Rev.} \textbf{\bibinfo{volume}{D86}},
  \bibinfo{pages}{075013} (\bibinfo{year}{2012}), \eprint{1207.1344}.

\bibitem[{\citenamefont{Espinosa
  et~al.}(2012{\natexlab{a}})\citenamefont{Espinosa, Grojean, Muhlleitner, and
  Trott}}]{grojean_higgs_1202}
\bibinfo{author}{\bibfnamefont{J.}~\bibnamefont{Espinosa}},
  \bibinfo{author}{\bibfnamefont{C.}~\bibnamefont{Grojean}},
  \bibinfo{author}{\bibfnamefont{M.}~\bibnamefont{Muhlleitner}},
  \bibnamefont{and} \bibinfo{author}{\bibfnamefont{M.}~\bibnamefont{Trott}},
  \bibinfo{journal}{JHEP} \textbf{\bibinfo{volume}{1205}}, \bibinfo{pages}{097}
  (\bibinfo{year}{2012}{\natexlab{a}}), \eprint{1202.3697}.

\bibitem[{\citenamefont{Espinosa
  et~al.}(2012{\natexlab{b}})\citenamefont{Espinosa, Grojean, Muhlleitner, and
  Trott}}]{grojean_higgs_1207}
\bibinfo{author}{\bibfnamefont{J.}~\bibnamefont{Espinosa}},
  \bibinfo{author}{\bibfnamefont{C.}~\bibnamefont{Grojean}},
  \bibinfo{author}{\bibfnamefont{M.}~\bibnamefont{Muhlleitner}},
  \bibnamefont{and} \bibinfo{author}{\bibfnamefont{M.}~\bibnamefont{Trott}},
  \bibinfo{journal}{JHEP} \textbf{\bibinfo{volume}{1212}}, \bibinfo{pages}{045}
  (\bibinfo{year}{2012}{\natexlab{b}}), \eprint{1207.1717}.

\bibitem[{\citenamefont{Passarino}(2013)}]{passarino_higgs_2012}
\bibinfo{author}{\bibfnamefont{G.}~\bibnamefont{Passarino}},
  \bibinfo{journal}{Nucl.Phys.} \textbf{\bibinfo{volume}{B868}},
  \bibinfo{pages}{416} (\bibinfo{year}{2013}), \eprint{1209.5538}.

\bibitem[{\citenamefont{Belanger et~al.}(2013)\citenamefont{Belanger, Dumont,
  Ellwanger, Gunion, and Kraml}}]{belanger_higgs_2012}
\bibinfo{author}{\bibfnamefont{G.}~\bibnamefont{Belanger}},
  \bibinfo{author}{\bibfnamefont{B.}~\bibnamefont{Dumont}},
  \bibinfo{author}{\bibfnamefont{U.}~\bibnamefont{Ellwanger}},
  \bibinfo{author}{\bibfnamefont{J.}~\bibnamefont{Gunion}}, \bibnamefont{and}
  \bibinfo{author}{\bibfnamefont{S.}~\bibnamefont{Kraml}},
  \bibinfo{journal}{JHEP} \textbf{\bibinfo{volume}{1302}}, \bibinfo{pages}{053}
  (\bibinfo{year}{2013}), \eprint{1212.5244}.

\bibitem[{atl()}]{atlas_comb_hcp_2012}
\bibinfo{note}{\url{http://cds.cern.ch/record/1499629/files/ATLAS-CONF-2012-170.pdf}}.

\bibitem[{cms()}]{cms_comb_hcp_2012}
\bibinfo{note}{\url{http://cds.cern.ch/record/1494149/files/HIG-12-045-pas.pdf}}.

\bibitem[{\citenamefont{Arkani-Hamed et~al.}(2002)\citenamefont{Arkani-Hamed,
  Cohen, Katz, and Nelson}}]{LH}
\bibinfo{author}{\bibfnamefont{N.}~\bibnamefont{Arkani-Hamed}},
  \bibinfo{author}{\bibfnamefont{A.}~\bibnamefont{Cohen}},
  \bibinfo{author}{\bibfnamefont{E.}~\bibnamefont{Katz}}, \bibnamefont{and}
  \bibinfo{author}{\bibfnamefont{A.}~\bibnamefont{Nelson}},
  \bibinfo{journal}{JHEP} \textbf{\bibinfo{volume}{0207}}, \bibinfo{pages}{034}
  (\bibinfo{year}{2002}), \eprint{hep-ph/0206021}.

\bibitem[{\citenamefont{Schmaltz}(2004)}]{SLH}
\bibinfo{author}{\bibfnamefont{M.}~\bibnamefont{Schmaltz}},
  \bibinfo{journal}{JHEP} \textbf{\bibinfo{volume}{0408}}, \bibinfo{pages}{056}
  (\bibinfo{year}{2004}), \eprint{hep-ph/0407143}.

\bibitem[{\citenamefont{Manohar and Wise}(2006)}]{octet}
\bibinfo{author}{\bibfnamefont{A.~V.} \bibnamefont{Manohar}} \bibnamefont{and}
  \bibinfo{author}{\bibfnamefont{M.~B.} \bibnamefont{Wise}},
  \bibinfo{journal}{Phys.Rev.} \textbf{\bibinfo{volume}{D74}},
  \bibinfo{pages}{035009} (\bibinfo{year}{2006}), \eprint{hep-ph/0606172}.

\bibitem[{\citenamefont{Appelquist et~al.}(2001)\citenamefont{Appelquist,
  Cheng, and Dobrescu}}]{UED}
\bibinfo{author}{\bibfnamefont{T.}~\bibnamefont{Appelquist}},
  \bibinfo{author}{\bibfnamefont{H.-C.} \bibnamefont{Cheng}}, \bibnamefont{and}
  \bibinfo{author}{\bibfnamefont{B.~A.} \bibnamefont{Dobrescu}},
  \bibinfo{journal}{Phys.Rev.} \textbf{\bibinfo{volume}{D64}},
  \bibinfo{pages}{035002} (\bibinfo{year}{2001}), \eprint{hep-ph/0012100}.

\bibitem[{\citenamefont{Cacciapaglia et~al.}(2010)\citenamefont{Cacciapaglia,
  Deandrea, and Llodra-Perez}}]{RPP}
\bibinfo{author}{\bibfnamefont{G.}~\bibnamefont{Cacciapaglia}},
  \bibinfo{author}{\bibfnamefont{A.}~\bibnamefont{Deandrea}}, \bibnamefont{and}
  \bibinfo{author}{\bibfnamefont{J.}~\bibnamefont{Llodra-Perez}},
  \bibinfo{journal}{JHEP} \textbf{\bibinfo{volume}{1003}}, \bibinfo{pages}{083}
  (\bibinfo{year}{2010}), \eprint{0907.4993}.

\bibitem[{\citenamefont{Cacciapaglia and Kubik}(2013)}]{RPPb}
\bibinfo{author}{\bibfnamefont{G.}~\bibnamefont{Cacciapaglia}}
  \bibnamefont{and} \bibinfo{author}{\bibfnamefont{B.}~\bibnamefont{Kubik}},
  \bibinfo{journal}{JHEP} \textbf{\bibinfo{volume}{1302}}, \bibinfo{pages}{052}
  (\bibinfo{year}{2013}), \eprint{1209.6556}.

\bibitem[{\citenamefont{Agashe et~al.}(2005)\citenamefont{Agashe, Contino, and
  Pomarol}}]{GHUwarped}
\bibinfo{author}{\bibfnamefont{K.}~\bibnamefont{Agashe}},
  \bibinfo{author}{\bibfnamefont{R.}~\bibnamefont{Contino}}, \bibnamefont{and}
  \bibinfo{author}{\bibfnamefont{A.}~\bibnamefont{Pomarol}},
  \bibinfo{journal}{Nucl.Phys.} \textbf{\bibinfo{volume}{B719}},
  \bibinfo{pages}{165} (\bibinfo{year}{2005}), \eprint{hep-ph/0412089}.

\bibitem[{\citenamefont{Cacciapaglia et~al.}(2009)\citenamefont{Cacciapaglia,
  Deandrea, and Llodra-Perez}}]{llodra_higgs_2009}
\bibinfo{author}{\bibfnamefont{G.}~\bibnamefont{Cacciapaglia}},
  \bibinfo{author}{\bibfnamefont{A.}~\bibnamefont{Deandrea}}, \bibnamefont{and}
  \bibinfo{author}{\bibfnamefont{J.}~\bibnamefont{Llodra-Perez}},
  \bibinfo{journal}{JHEP} \textbf{\bibinfo{volume}{0906}}, \bibinfo{pages}{054}
  (\bibinfo{year}{2009}), \eprint{0901.0927}.

\end{thebibliography}

\end{document}